\documentclass{article}

\usepackage{arxiv}
\usepackage{graphicx}
\usepackage[utf8]{inputenc} 
\usepackage[T1]{fontenc}    
\usepackage{hyperref}       
\usepackage{url}            
\usepackage{booktabs}       
\usepackage{amsfonts}       
\usepackage{nicefrac}       
\usepackage{microtype}      
\usepackage{lipsum}
\usepackage{textgreek}
\title{Next Generation of Star Patterns}

\author{
  Hadi Mansourifar\\
  Department of Computer Science\\
  University of Houston\\
  Houston, Texas \\
  \texttt{hmansourifar@uh.edu} \\
   \And
 Weidong Shi \\
  Department of Computer Science\\
  University of Houston\\
  Houston, Texas \\
  \texttt{wshi3@central.uh.edu} \\
}

\begin{document}
\maketitle
\begin{abstract}
In this paper we present two new ideas for generating star patterns and filling the gaps during the tile operation. Firstly, we introduce a novel parametric method based on concentric circles for generating stars and rosettes. Using proposed method, completely different stars and rosettes and a set of new and complex star patterns convert to each other only by changing nine parameters. Secondly, we demonstrate how three equal tangent circles can be used as a base for generating tile elements. For this reason a surrounded circle is created among tangent circles, which represents the gaps in hexagonal packing. Afterwards, we use our first idea for filling the tangent circles and surrounded circle. This parametric approach can be used for generating infinite new star patterns, which some of them will be presented in result section.Two Android apps of proposed method called Starking and Tilerking are available in Google app store.

\end{abstract}


\section{Introduction}
Traditional symbols have significant role in cultural memory of each society. Utilizing traditional symbols in new intellectual products can help cultural continuity from one generation to the next. Moreover new intellectual products can be inspired by old patterns and can produce own symbols. For instance star patterns are one of the traditional symbols which can be used in social contexts as a reference to the past. However, in the field of drawing new types of star patterns, main problem is to find a regular approach for filling the gaps. Without this regular approach, tiling methods cannot be combined with pattern generation techniques and gaps should be filled by inference algorithm. Filling the gaps by an inference algorithm is mostly used by polygon in contact method \cite{12}, \cite{13}, \cite{14}. Therefore polygon in contact method cannot be combined with pattern generation techniques for generating new types of Islamic stars. In this paper, we propose two new parametric methods for filling the gaps in hexagonal packing. In the first method, a hidden relationship between concentric circles is parameterized in order to generate stars and rosettes. Using this new parametric approach, completely different stars and rosettes convert to each other only by changing nine parameters. Moreover proposed method has sufficient flexibility for producing new generation of star patterns. In the second method we use three equal tangent circles as a base of tile operation. Then a surrounded circle is recognized among three tangent circles which represents the gaps during the tile operation. After finding the properties of surrounded circle, our parametric method is used for filling the tangent circles and surrounded circle. Finally this triplet pattern is repeated to cover the 2D space. This parametric approach for generating tile elements, provides the designers with a powerful tool to produce new types of tile patterns. The rest of this paper is organized as follows: section 2 reviews some of the previous works in the field of Islamic stars. Section 3 demonstrates main idea of the paper. Section 4 presents required parameters, formulas and algorithms for generating Islamic stars by concentric circles. Section 5 demonstrates a new idea for generating tile elements by tangent circles. Section 6 presents some of our exclusive results. Finally, Section 7 concludes the paper and provides required discussion and possible future directions.

\begin{figure}[ht!]
\centering
\includegraphics[width=90mm]{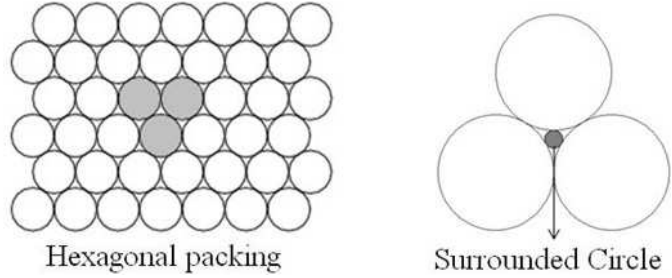}
\caption{ Three tangent circles and surrounded circle \label{overflow}}
\end{figure}
\section{	Background}
\label{sec:others}
The most researches about Islamic stars fall in two main categories. First, symmetry as a traditional approach has devoted many researches to itself. Researches about symmetry are not restricted to Islamic stars \cite{9},\cite{20}. Also, Alexander \cite{2} developed a FORTRAN program for generating the 17 types of design in the Euclidean plane. But exclusively in the field of Islamic stars, Grünbaum and Shephard \cite{11} provided a complete set of mathematical tools for decomposing periodic Islamic patterns by their symmetry groups. Abas and Salman \cite{1} applied this symmetry groups on a wide-ranging of historical patterns. Then Ostromoukhov \cite{20} proposed a mathematical tool for computer generated ornamental patterns. In the other researches in this field Dewdney \cite{7} presented a new arrangement based on reflecting lines through a regular classification of circles. Castéra \cite{5} presented a method based on the creation of skeleton of eightfold stars and safts. Aljamali and Banissi \cite{3} proposed a rational classification of Islamic Geometric Patterns (IGP) based on the Minimum Number of Grids (MNG) and Lowest Geometric Shape (LGS), which used in the construction of the symmetric elements. Despite the power of techniques that are based on symmetry, they are not able to propose a flexible approach for generating new star patterns. Dunham in his valuable research \cite{8} adjusted several Islamic patterns to the hyperbolic plane. Second category of researches about Islamic stars was established by Hankin \cite{12},\cite{13},\cite{14}. In recent years, scholars such as Lee \cite{18}, Critchlow \cite{6}, Craig S. Kaplan \cite{16}, \cite{15} and Craig S. Kaplan and David H. Salesin \cite{17} have referred to Hankins’s polygons-in contact method. Finally Kaplan developed an applet for generating decorative patterns. Main problem of polygons-in contact method is filling the gaps by an inference algorithm. Bastanfard and Mansourifar \cite{4} purposed a new method for generating stars and filling the gaps. But all these methods cannot present a common approach for generating stars and rosettes. On the other hand previous methods can generate a restricted range of stars or rosettes and results are predictable. Compared with all the previous works our approach presents a novel method which is efficient in generating new generation of star patterns.

\section{Generating Star Patterns by Concentric Circles}
In this section we demonstrate a hidden relationship between concentric circles, which can be used for generating stars and rosettes.  For analyzing given  star pattern in part (a) of figure 2, the lines are hidden and beginning and end of the lines are clarified by marked points as shown in Part (b) of figure 2. As illustrated in part (c) of figure 2 all the marked points are located on three concentric circles. In fact each concentric circle is divided by eight marked points to eight equivalent segments.
\begin{figure}[ht!]
\centering
\includegraphics[width=90mm]{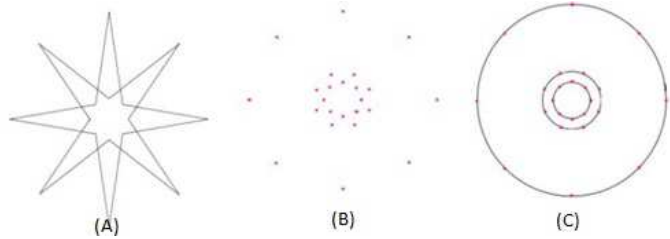}
\caption{ The role of concentric circles to generate star patterns \label{overflow}}
\end{figure}
After recognizing concentric circles on stars, a specific number is assigned to each marked point as shown in figure 3. For this reason various start angles is selected for denoting first marked point on each concentric circle. For example first marked point of first concentric circle is located on 0 degree. But first marked point of second concentric circle is located on 22.5 degree and first marked point of third concentric circle is located on 45 degree.
\begin{figure}[ht!]
\centering
\includegraphics[width=70mm]{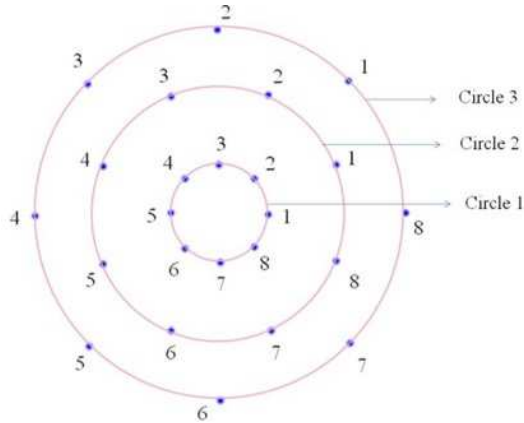}
\caption{ Dividing concentric circles
\label{overflow}}
\end{figure}
In this case the angle difference between first marked point of a concentric circle and first marked point of next concentric circle is 22.5 degree. In general, the angle difference between first marked points of two adjacent concentric circles is calculated as follows.
\[\Delta = 360/ (2N)\] Where, number of segments on each circle is denoted by N. For generating a star pattern, marked points of two adjacent concentric circles should be connected according to a specific regularity. The instruction for connecting  marked points to each other is described in figure 4. For example, marked points 4 and 5 from second circle have been connected to marked point 4 of third circle. In fact marked point (x) and marked point ((x+1) mod (N)) from first circle have been connected to marked point (x) from the next circle. Regardless of the number of concentric circles this regularity always exists between two adjacent concentric circles.

\begin{figure}[ht!]
\centering
\includegraphics[width=65mm]{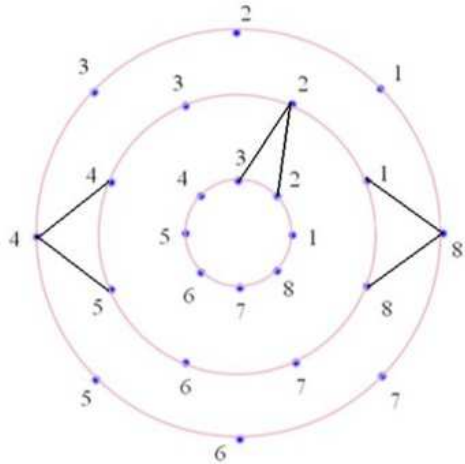}
\caption{ Using numbers for connecting marked points.
\label{overflow}}
\end{figure}

\begin{figure}[ht!]
\centering
\includegraphics[width=90mm]{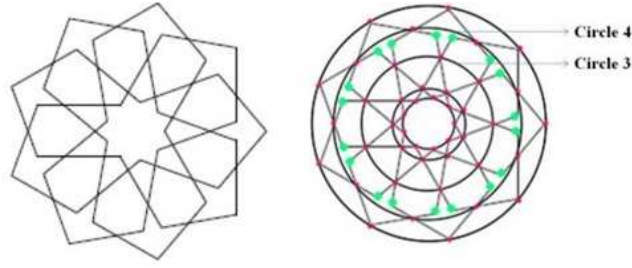}
\caption{ Difference between stars and rosettes
\label{overflow}}
\end{figure}

\subsection{Concentric Circles for Generating Rosettes}
Even though stars and rosettes seem different, they are strikingly similar in many ways.For instance, concentric circles can be used for drawing rosettes, too.  Figure 5 (left) shows an 8-pointed rosette. In this rosette four concentric circles can be recognized as shown in right image of figure 5.These circles and their marked points follow from stars order.But there are points that do not follow from stars order. These special points have been located between $circle (3)$ and $circle (4)$ and are shown by green color. Therefore regularity of these points should be found between $circle (3)$ and $circle (4)$. Furthermore special points are located on the specific circle. Radius of this specific circle is bigger than $circle (3)$ and smaller than $circle (4)$. The name of this specific circle is Green circle. Therefore Green circle is defined as a circle that, special points are located on it.To obtain the position of each special point, radius of Green circle and angle of each special point are required. For calculating the radius of Green circle, the $spr$ parameter is decreased from radius of $circle (4)$. Hence, angle of each special point can be calculated, using the angles of $circle (4)$. As illustrated in figure 6 two special points are recognized for every marked point on $circle (4)$. For instance marked point (2) of $circle (4)$ is connected to two special points, which have been shown by number 2.
\begin{figure}[ht!]
\centering
\includegraphics[width=70mm]{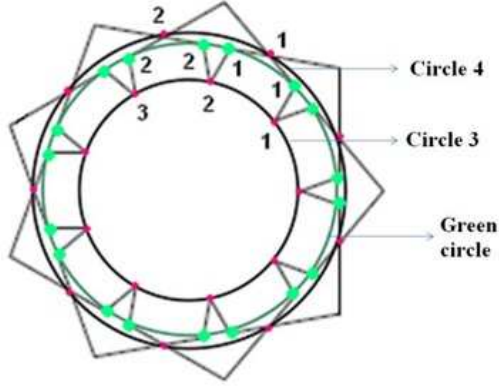}
\caption{ Rosettes formulation.
\label{overflow}}
\end{figure}
One special point’s angle is $\alpha$ degree smaller than $angle (2)$ of $circle (4)$ and next special point’s angle is $\alpha$ degree bigger than $angle (2)$ of $circle (4)$. Therefore using the $spr$ and parameters, position of special points is obtained as follows.

\[ xsp1=(r(w)- spr) * cos(angle (i) - \alpha + x0 \]
\[ ysp1=(r(w)- spr) * sin(angle (i) - \alpha + y0 \]
\[ xsp2=(r(w)- spr) * cos(angle (i) + \alpha + x0 \]
\[ ysp2=(r(w)- spr) * sin(angle (i) + \alpha + y0 \]

Where $(xsp1,ysp1)$ is position of first special point of $angle (i)$ from $circle (w)$ and $(xsp2,ysp2)$ is position of second special point of $angle (i)$ from $circle (w)$. Special points are located between every two possible concentric circles. In these situations the bigger circle is a special circle which is the only difference between stars and rosettes.

\section{Algorithm}

In this section we introduce required parameters and algorithms for drawing stars and rosettes by concentric circles. For parametrizing the stars and rosettes, number of concentric circles is denoted by $S$ parameter and circle division number is specified by $N$ parameter. Also $angle (i)$ denotes angles of each concentric circle, $circle (w)$ and $r (w)$ denote circles and their radiuses.
\subsection{Algorithm of Dividing Circles}
For drawing stars, concentric circles should be divided to a set of angles. Then a valid amount should be assigned to each angle. Following algorithm presents required algorithm.\newline
\textbf{1}.	Set t =0 and valid amount to N and S \newline
\textbf{2}.	Set i =1\newline
\textbf{3}.	Set j =1\newline
\textbf{4}.	angle ( j ) of circle (i) = t\newline
\textbf{5}.	t = t + (360 / N) \\
\textbf{6}.	j = j +1 ; if  j < =N go to step 4\newline
\textbf{7}.	t = t + (180 / N)\newline
\textbf{8}.	i = i +1  if i < =S go to \quad step 3\newline
\subsection{Algorithm of Drawing Stars and Rosettes}
In this section a common algorithm is presented for drawing the stars and rosettes. Using this algorithm various stars and rosettes are converted to each other only by changing nine parameters.

\textbf{1}.\quad	Set valid amounts for $S$, $N$, $w$, $r (1)$ to $r(S)$\newline
\textbf{2}.\quad	$w =1$\newline
\textbf{3}.\quad	$i=1$\newline
\textbf{4}.\quad If $circle ( w+1)$ != special circle then \newline
$gx1=r(w) * cos(angle(i)) +x0 $\newline
$gy1=r(w) * sin(angle(i)) +y0 $\newline
$gx2=r(w) * cos(angle(i+1) \quad mod \quad N) +x0$ \newline
$gy2=r(w) * sin(angle(i+1) \quad mod \quad N) +y0$ \newline
$gx3=r(w+1) * cos(angle(i)) +x0$ \newline
$gy3=r(w+1) * sin(angle(i)) +y0 $\newline
\textbf{5}.\quad Draw lines From $(gx1, gy1)$ to $(gx3, gy3)$\newline
From $(gx2, gy2)$ to $(gx3, gy3)$\newline
\textbf{6}. \quad  $i=i+1 \newline
if  \quad( $i<=N$) \quad go \quad to \quad step \quad 4 \newline
Else\newline
\textbf{7}.\newline
gx1=r(w) * cos(angle(i)) +x0\newline
gy1=r(w) * sin(angle(i)) +y0 \newline
gx2=r(w) * cos(angle(i+1) mod N) +x0 \newline
gy2=r(w) * sin(angle(i+1) mod N) +y0 \newline
gx3=r(w+1)-spr * cos(angle(i-\textalpha)) +x0 \newline
gy3=r(w+1)-spr * sin(angle(i-\textalpha)) +y0 \newline
gx4=r(w+1)-spr * cos(angle(i+\textalpha)) +x0 \newline
gy4=r(w+1)-spr * sin(angle(i+\textalpha)) +y0 \newline
gx5=r(w+1) * cos(angle(i)) +x0 \newline
gy5=r(w+1) * sin(angle(i)) +y0 \newline
Draw lines: \quad From \quad (gx1, gy1)$ \quad to \quad$(gx3 , gy3) \quad and \quad From \quad (gx2, gy2) \quad to \quad (gx3, gy3) \newline
\quad From \quad(gx3, gy3) \quad to \quad (gx5, gy5) \newline \quad From \quad (gx4, gy4) \quad to \quad (gx5, gy5)\newline
\textbf{8}.\quad $i= i +1$ \quad if \quad $(i <=N) \quad go \quad to \quad step \quad 7\newline
\textbf{9}.\quad $w = w + 1$ \quad if \quad $(w< S)$ \quad go \quad to \quad step 3\newline

Here we present two examples based on our method. In these
examples two completely different stars and rosettes convert to each other only by changing nine parameters. First we set parameters according to part 1 of table 1.
\begin{table}[h!]
\centering
 \begin{tabular}{||c c c c c c c c c c||} 
 \hline
 Part & N & \textalpha & r1 & r2 & r3 & r4 & Spr & S &  Sp \\ [0.5ex] 
 \hline\hline
 1 & 8 & 0 & 51 & 70 & 172& --& --& 3&-- \\ 
 \hline
 2 & 9 & 48 & 93 & 225 & --& 180 & -68  & 2 & 2 \\
 
 \hline
\end{tabular}
\bigskip
\caption{Required parameters to generate stars of Figure7.}
\label{table:1}
\end{table}
Part 1 of figure 7 shows the result of this example. In another example we set parameters according to part 2 of table 2. Part 2 of figure 7 shows the result of example.
\bigskip
\bigskip
\begin{figure}[ht!]
\centering
\includegraphics[width=90mm]{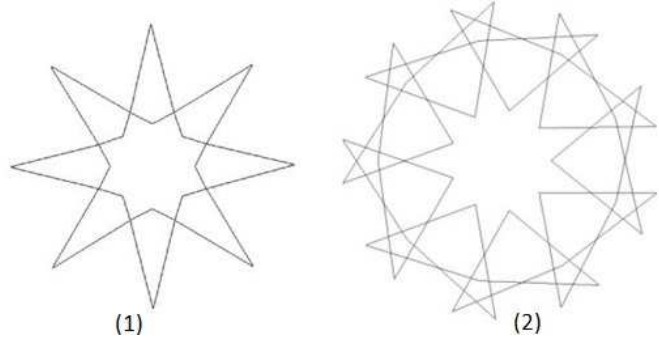}
\caption{ Different patterns convert to each other by changing 9 parameters.
\label{overflow}}
\end{figure}
\subsection{Generating Desired Sketches}
In the field of Islamic stars two points should be considered as follows. First it’s important for the sides of the outer hexagons in a rosette to be parallel. Second it's important to be able to control the angle formed by the rays meeting at the outermost circle. Figure 8 shows a desired 8-pointed rosette and its concentric circles. In this rosette, gradient of straight line between $angle (1)$ of $circle (1)$ and $angle (1)$ of $circle (2)$ is equal to gradient of straight line between $angle (1)$ of $circle (2)$ and $angle (1)$ of $circle (3)$. In general, gradient of straight line between $angle (1)$ of $circle (w)$ and $angle (1)$ of $circle (w+1)$ is always equal to gradient of straight line between $angle (1)$ of $circle (1)$ and $angle (1)$ of circle (2). For generating such desired sketches radius of concentric circles are assigned to appropriate amounts as follows.

\[ r(w)= (\alpha *x_m- y_m)/(\theta *cos(\alpha_ n ) - sin(\alpha_ n )) \]
\begin{figure}[ht!]
\centering
\includegraphics[width=40mm]{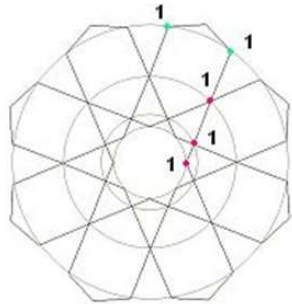}
\caption{ Analyzing desired rosette.
\label{overflow}}
\end{figure}
\section{Tile operation}
In this section we demonstrate our parametric approach for generating tile elements. For this reason, four steps should be taken as follows.\\
\textbf{1}. Drawing three equal tangent circles as a primitive sketch of tile element.\\
\textbf{2}. Recognizing a surrounded circle among tangent circles and calculating its properties.\\
\textbf{3}. Filling the tangent circles and surrounded circle by proposed method of section 3 and 4.\\
\textbf{4}. Repeating filled tangent circles and their surrounded circle to cover the space.\\
 Figure 9 (left) describes three tangent circles which gap area among them has been clarified by gray color. The right part of figure 9 shows that, the gap area can be bounded by a surrounded circle.
\bigskip
\bigskip
\begin{figure}[ht!]
\centering
\includegraphics[width=120mm]{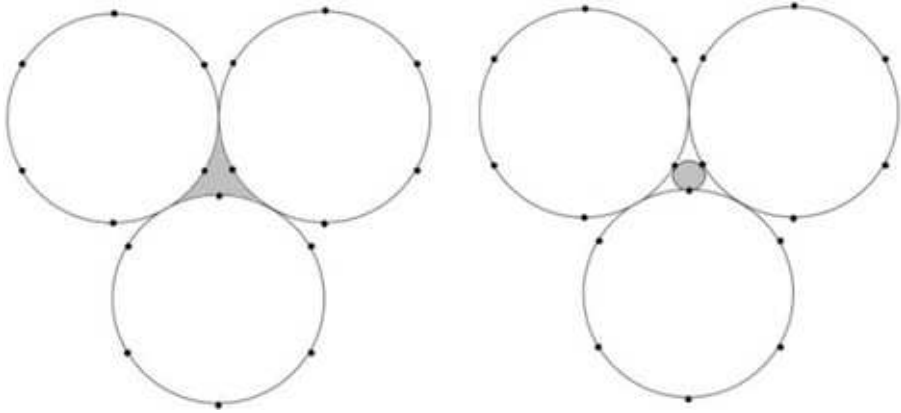}
\caption{ Bounding the gap area by surrounded circle.
\label{overflow}}
\end{figure}
\subsection{Surrounded circle}
Surrounded circle has two properties: the position of origin and radius. These properties is used for filling the surrounded circle by concentric circles.
\begin{figure}[ht!]
\centering
\includegraphics[width=70mm]{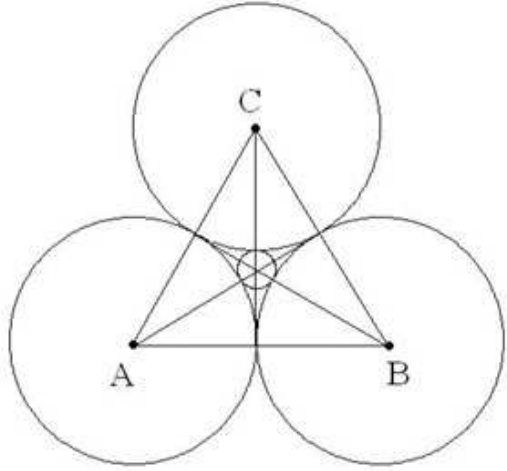}
\caption{ Calculating properties of surrounded circle.
\label{overflow}}
\end{figure}
The origin of surrounded circle is located in center of equilateral triangle as shown in \quad figure 10. Therefore the origin of surrounded circle can be calculated as follows.\bigskip
\[S_x=(x_A+ x_B)\]   \[ S_y=y_A+ (y_C - y_A)/3 \]  \\ Where ($S_x$,$S_y$) is the origin of surrounded circle and radius of tangent circles is calculated as follows.\[ M=S_y-y_A  \]\[ Sin (30) =\frac{M}{(R+r)}  \] \[ r=2M-R \] 
\section{Results}
Using the parametric approach, unpredictable results can be generated. In this section we introduce some of the new star patterns which have been generated exclusively by our method. Figure 11 shows two new star patterns which can be converted to each other only by changing nine parameters. Required parameters for generating these patterns are presented in table 2. Moreover Figures 12,13 and 14 show three new tile patterns which can be converted to each other only by changing nine parameters, too. Required parameters for filling tangent circles are presented in table 3. On the other hand for filling the surrounded circle two concentric circles have been used. These concentric circles are as follows: outer circle and inner circle. Radius of outer circle is equal to radius of surrounded circle and inner radius is higher or equal to $0.5* (outer \quad radius)$.Note that, two Android apps of proposed method called Starking \cite{22} and Tilerking \cite{23} have been published in Google app store and they are available in both free and pro versions.
\begin{figure}[ht!]
\centering
\includegraphics[width=120mm]{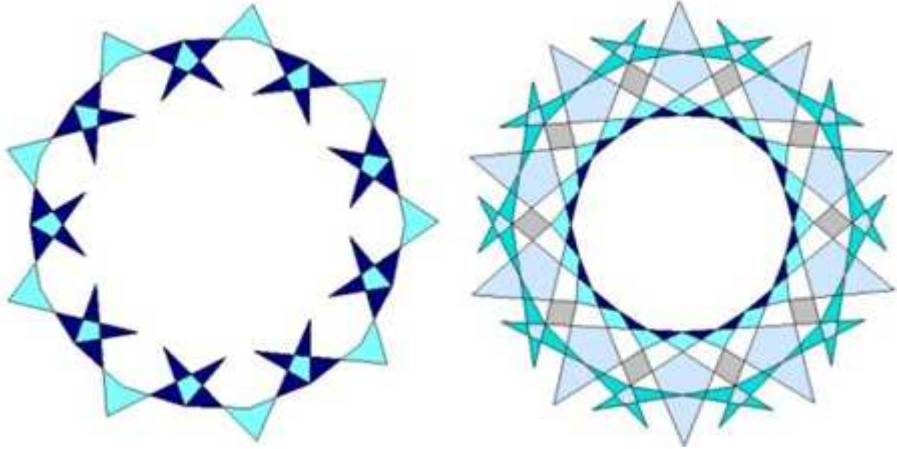}
\caption{Star results: New types of star patterns generated by proposed method.
\label{overflow}}
\end{figure}
\begin{table}[h!]
\centering
 \begin{tabular}{||c c c c c c c c c c||} 
 \hline
 Result & N & \textalpha & r1 & r2 & r3 & r4 & Spr & S &  Special circle \\ [0.5ex] 
 \hline\hline
 Left & 9 & 34 & 191 & 189 & 226 & --& 89& 3&3 \\ 
 \hline
 Right & 10 & 62 & 172 & 109 & 133& 125 & -100  & 4 & 2 \\
 
 \hline
\end{tabular}
\bigskip
\caption{Required parameters for each star results.}
\label{table:1}
\end{table}
\begin{figure}[ht!]
\centering
\includegraphics[width=90mm]{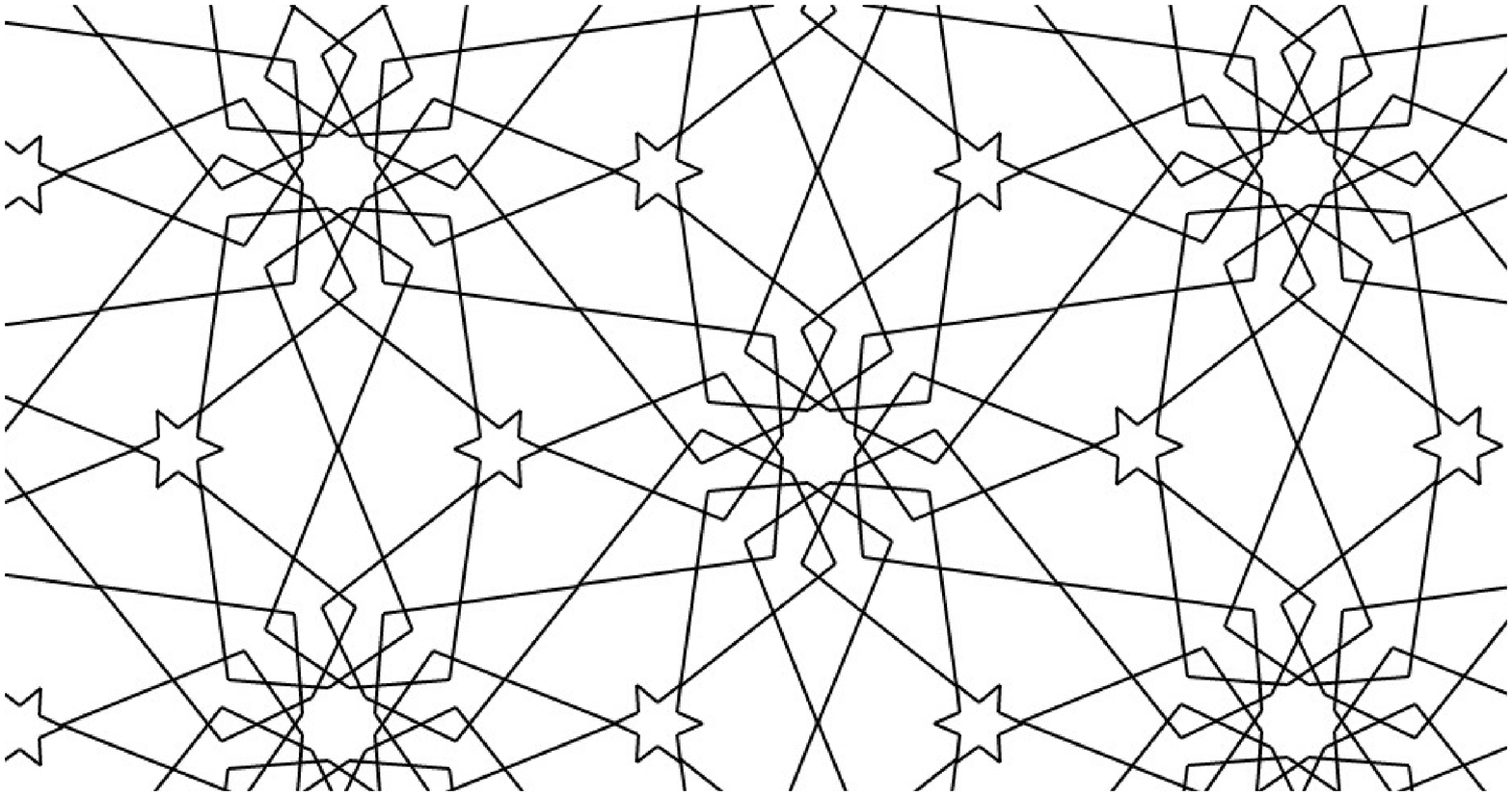}
\caption{ Tiling Result 1.
\label{overflow}}
\end{figure}
\begin{figure}[ht!]
\centering
\includegraphics[width=90mm]{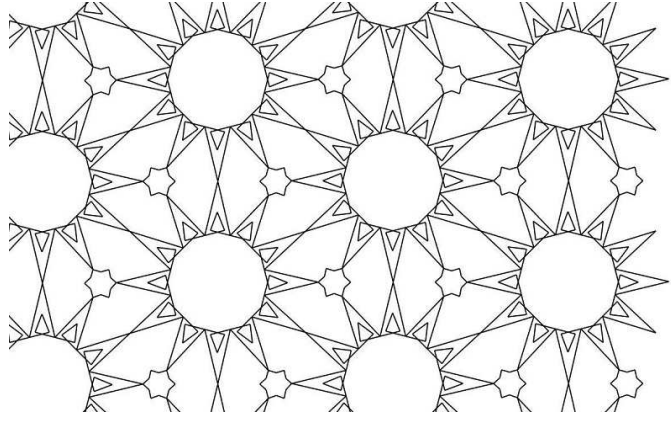}
\caption{ Tiling Result 2.
\label{overflow}}
\end{figure}
\begin{table}[h!]
\centering
 \begin{tabular}{||c c c c c c c c c c||} 
 \hline
 Result & N & \textalpha & r1 & r2 & r3 & r4 & Spr & S &  Special circle \\ [0.5ex] 
 \hline\hline
 1 & 12&	53&	171&	23	&214&	--&	-50&	2&	2 \\ 
 \hline
 2 & 6	&10&	143&	145&	179&	--&-70	&4	&2 \\
 \hline
 3 & 12	&23&	123	&85	&178& --&	-7&	3	&2 \\
 \hline
\end{tabular}
\bigskip
\caption{Required parameters for filling tangent circles.}
\label{table:1}
\end{table}
\bigskip
\bigskip
\bigskip
\bigskip
\bigskip
\bigskip
\bigskip
\bigskip
\begin{figure}[ht!]
\centering
\includegraphics[width=90mm]{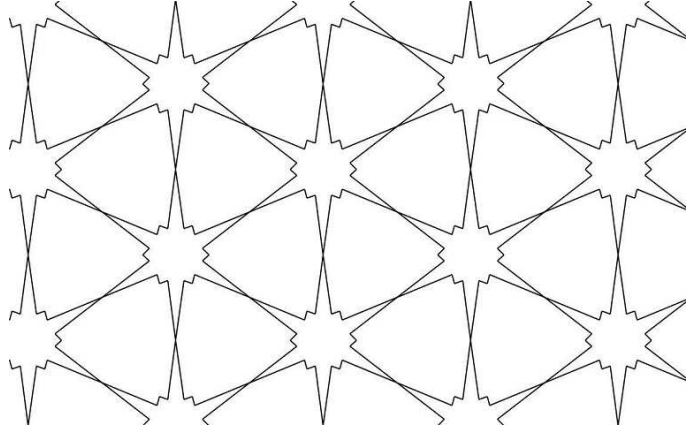}
\caption{ Tiling Result 3.
\label{overflow}}
\end{figure}
\section{Conclusion}
In this paper, we have purposed two new ideas for generating star patterns and filling the gaps during the tile operation. New ideas are as follows.
\begin{itemize}
  \item Utilizing the concentric circles for parametrizing stars and rosettes.
  \item Utilizing three tangent circles as a base for drawing tile elements.
\end{itemize}
These new approaches have many benefits. First, proposed method presents a common approach for drawing stars and rosettes while other methods have not such capability. Second, only by changing nine parameters, star patterns from the simplest model to the most complex one can be converted to each other easily. Third, despite the symmetry based methods, our method uses a parametric approach for generating tile elements. For this reason we extract tile element from three tangent circles. Therefore various tile elements can be converted to each other easily. Furthermore using the purposed method, unpredictable stars and rosettes are generated. Some of these stars or rosettes cannot easily be produced by other methods. On the other hand our method uses a parametric approach for filling the gaps. Therefore inside sketch of gap area can be controlled with a parametric approach. The time complexity of purposed algorithm is O $(N \times S)$. Where $N$ is number of segments on each concentric circle and $S$ is number of concentric circles.

\bibliographystyle{unsrt}

\begin{thebibliography}{1}

\bibitem{kour2014real}
George Kour and Raid Saabne.
\newblock Real-time segmentation of on-line handwritten arabic script.
\newblock In {\em Frontiers in Handwriting Recognition (ICFHR), 2014 14th
  International Conference on}, pages 417--422. IEEE, 2014.

\bibitem{kour2014fast}
George Kour and Raid Saabne.
\newblock Fast classification of handwritten on-line arabic characters.
\newblock In {\em Soft Computing and Pattern Recognition (SoCPaR), 2014 6th
  International Conference of}, pages 312--318. IEEE, 2014.

\bibitem{hadash2018estimate}
Guy Hadash, Einat Kermany, Boaz Carmeli, Ofer Lavi, George Kour, and Alon
  Jacovi.
\newblock Estimate and replace: A novel approach to integrating deep neural
  networks with existing applications.
\newblock {\em arXiv preprint arXiv:1804.09028}, 2018.

\end{thebibliography}


\begin{thebibliography}{1}
\bibitem{1} 
Abas, S. and Salman, A.  Geometric and group-theoretic methods for computer graphics studies of Islamic symmetric Patterns. 
. 
\textit{The \LaTeX\ Companion}. 
Comput. Graph. For. 11, 1, 43–53,1992.
 
\bibitem{2} 
Howard Alexander,The computer/plotter and the 17 ornamental design types.In Proceedings of SIGGRAPH 1975.
 

\bibitem{3} 
Ahmad M. Aljamali and Ebad Banissi Geometric modeling: techniques, applications, systems and tools 2004.R. Nicole,Title of paper with only first word capitalized, J. Name Stand. Abbrev.
\bibitem{4} 
Bastanfard and Mansourifar, A novel decorative Islamic star Pattern generation Algorithm, In IEEE Proceedings of ICCSA 2010 proceedings, page 111-117, Japan, March 2010.
\bibitem{5} 
Caste'ra, J.-M. 1999. Arabesques: Decorative Art in Morocco. ACR Edition
\bibitem{6} 
Keith Critchlow. Islamic Patterns: An Analytical and Cosmological Approach. Thames and Hudson, 1976.
\bibitem{7} 
Dewdney, A. 1993, The Tinkertoy Computer and Other Machinations. W. H. Freeman, 222–230.
\bibitem{8} 
Dunham. D. 2001. Hyperbolic Islamic patterns—A beginning. In Bridges 2001 Proceedings, R. Sarhangi, Ed.
\bibitem{9} 
Andrew Glassner. Frieze groups. IEEE Computer Graphics and Applications,16(3):78–83, 1996.
\bibitem{10} 
Grünbam, and Shephard, G. C. 1987. Tilings and Patterns. W. H. Freeman.
\bibitem{11} 
Grünbam, and Shephard, G. C. 1992. Interlace patterns in Islamic and Moorish art. Leonardo 25, 331–339.
\bibitem{12} 
E. H. Hankin. The Drawing of Geometric Patterns in Saracenic Art, volume 15 of Memoirs of the Archaeological Society of India,1925.
\bibitem{13} 
E. H. Hankin. Examples of methods of drawing geometrical arabesque patterns. The Mathematical Gazette, pages 371–373,  1925.
\bibitem{14} 
E. Hanbury Hankin. Some difficult Saracenic designs
III.The Mathematical Gazette, pages 318–319, December 1936.
\bibitem{15} 
Craig  S.  Kaplan.  Computer  generated  Islamic  star patterns. In Reza Sarhangi, editor, Bridges 2000 Proceedings.
	

\bibitem{16} 
Craig S. Kaplan. Islamic Star Patterns from Polygons in Contact. Proceedings of Graphics Interface 2005,177-185
\bibitem{17} 
Craig  S.  Kaplan  and  David  H.  Salesin.  Islamic  star patterns in absolute geometry. ACM Trans. Graph.,23(2):97–119, 2004.

\bibitem{18} 
A.J. Lee. Islamic star patterns. Muqarnas, 4:182–197, 1987.
\bibitem{19} 
G. E. Martin, Transformation Geometry. An Introduction toSymmetry, Springer, New York (1982).
\bibitem{20} 
Ostromoukhov, V. 1998. Mathematical tools for computer-generated ornamental patterns. In Electronic Publishing, Artistic Imaging and Digital Typography. In Lecture Notes in Computer Science, vol. 1375. Springer-Verlag, New York, 193–223.
\bibitem{21} 
Taprats.http://www.cgl.uwaterloo.ca/~csk/washington
/taprats, last seen (Sep 2018).
\bibitem{22}
https://play.google.com/store/apps/details? id=com.hexappon.starkingfree . last seen(Sep 2018)

\bibitem{23}
https://play.google.com/store/apps/details?id=com.hexappon.TilerKingFree. last seen(Sep 2018)
\end{thebibliography}

\end{document}